# Roles of Diffusion Dynamics and Molecular Concentration Gradients in Cellular Differentiation and Three-Dimensional Tissue Development

Richard J. McMurtrey, MD, MSc

Institute of Neural Regeneration & Tissue Engineering, Highland, UT, United States
Email: *richard.mcmurtrey@neuralregeneration.org*

*ABSTRACT*

Recent advancements in the ability to construct three-dimensional (3D) tissues and organoids from stem cells and biomaterials have not only opened abundant new research avenues in disease modeling and regenerative medicine but also have ignited investigation into important aspects of molecular diffusion in 3D cellular architectures. This paper describes fundamental mechanics of diffusion with equations for modeling these dynamic processes under a variety of scenarios in 3D cellular tissue constructs. The effects of these diffusion processes and resultant concentration gradients are described in the context of the major molecular signaling pathways in stem cells that both mediate and are influenced by gas and nutrient concentrations, including how diffusion phenomena can affect stem cell state, cell differentiation, and metabolic states of the cell. The application of these diffusion models and pathways is of vital importance for future studies of developmental processes, disease modeling, and tissue regeneration.

## 1. INTRODUCTION

Life exists at the interface of numerous molecular, chemical, and physical processes, and as part of this precarious balance, life must both exploit and overcome various features of these phenomena. One of the most important physical processes that both defines and limits cellular functions in the human body is that of diffusion. Evidence suggests that diffusion of molecular factors may play vital roles in the self-organization of tissue architecture and determination of cellular identity in development, including factors affecting potency, differentiation, metabolic state, and functions of cells and tissues. Many of these developmental and metabolic processes and signaling pathways remain to be studied and elucidated, but the ability to mathematically model the role of diffusion processes with precise theoretical determinations opens a valuable and expansive field of mathematical study in stem cell biology, developmental biology, tissue engineering, and disease modeling.

Recent major advancements in biomaterials and stem cell culture have enabled the construction of complex three-dimensional (3D) multi-cellular organoid tissues. In some cases, cells are able to self-organize within a homogenous biomaterial scaffold [1], in some cases the cellular architecture may be guided by more complex configurations of patterned topographical and biochemical cues within the construct [2], while in other cases cells may be grown into aggregate multi-cellular spheroids without the addition of biomaterials [3]. These organoid constructs enable incredible new capabilities for researching numerous biological processes in a controlled *in vitro* environment, including the study of organ development, stem cell growth and differentiation, and cell signaling factors involved in the formation of cellular identity and spatial patterning. Importantly, these organoid technologies also open up a vast number of clinical applications, including disease modeling, pharmacological and toxicological drug testing, tumor models that direct personalized chemotherapy, and tissue reconstruction for regenerative medicine. However, the formation of 3D cellular cultures also gives rise to new complexities in physical diffusion phenomena that are not present in more traditional two-dimensional (2D) culture systems and which warrant detailed examination.

*2. INTERACTIONS OF DIFFUSION PHENOMENA WITH STEM CELL FUNCTION AND TISSUE DEVELOPMENT*

*2.1 The Emergence of Diffusion Phenomena in 3D Tissue Constructs*

With the formation of conglomerate cell cultures and engineered tissues, new diffusion dynamics arise in the 3D constructs. Delivery of nutrients like oxygen, glucose, fats, and amino acids to cells in such constructs is effectively more limited, which can critically affect *in vitro* tissue development as well as integration of the construct into the body after implantation. Diffusion-limited growth and inadequate mass transport of nutrients or signaling factors into the deeper or more sequestered regions of the construct tends to decrease cell survival and tissue size, and metabolism of the diffusant further decreases its availability and alters its spatial concentration profile through the construct [4]. The shape of the concentration gradient through a tissue is affected by the local conditions of the system, which is important because the shape of this gradient may produce differential downstream consequences on the development of stem cells based on their positioning within tissues. In addition, concentration gradients are also known to play important roles in axonal guidance, although much remains to be explored on the mechanisms of how such gradients are established at the proper place and time in tissues and how multiple gradients interact with each other to influence developing cells and tissue architecture. Thus modeling these mechanisms helps to understand both normal development and may also relate to previously unknown mechanisms of certain developmental pathologies.

In general, there is a Gaussian-shaped curve through unbound space for limited and unmetabolized substances diffusing through a homogenous medium of any dimensionality, while an unmetabolized substance in constant or unlimited supply in tissue constructs will generally produce the shape of a complementary error function (erfc) when diffusing primarily in one dimension, the shape of a Bessel function of the first kind (order zero) when diffusing primarily in two dimensions, and the shape of a hyperbolic curve when diffusing in three dimensions [4]. The introduction of constant (zero order) metabolism of the diffusant generally results in a parabolic concentration curve in any dimensionality [4]. Multiple different molecular factors can simultaneously overlap with entirely different concentration gradients depending on the conditions and characteristics of both the cellular tissue and each molecular factor. The initial concentration of diffusant at the tissue interface ($C_o$) will proportionally influence the concentration values throughout the tissue, while the diffusion coefficient ($D$) has a more complex role, proportionally relating the molar flux of diffusant to the spatial concentration gradient, which, in essence, ultimately describes how easily a particular diffusant moves through a medium. Diffusion limitations can be partially overcome through a variety of methods, which include the general approaches of increasing nutrient concentrations in surrounding fluid, decreasing the diffusion coefficient in the construct material (i.e., increasing permeability to nutrients), decreasing the diffusion range or depth of the tissue construct, increasing convective flow or perfusion of nutrient, or decreasing nutrient consumption, and each of these approaches has consequences for the tissue construct.

Importantly, diffusion limitations can also be desirable, producing concentration gradients into or out of the tissue construct that mediate essential developmental processes and activate or inhibit vital cell signaling cues. It is known that diffusion gradients of numerous morphogenic signaling factors play extensive roles in the differentiation and architectural formation of neural tissue, including, for example, regional gradients of sonic hedgehog (SHH), wnt protein (WNT), bone morphogenic protein (BMP), fibroblast growth factor (FGF), retinoic acid (RA), and reelin (RELN) [5]. The limited diffusion capacity of biomaterials not only causes decreased concentration of externally-supplied nutrients within the construct, but also results in increased internal concentrations of endogenous factors secreted by cells. This property of biomaterials enables endogenous signaling factors to form local regions of concentration gradients similar to what occurs in endogenous developing tissue, and this property likely provides essential self-organization capabilities of cells in 3D organoid constructs over standard 2D cultures under the same culture conditions. Thus, several analytic models are presented herein, including equations for both inward and outward diffusion, to describe diffusant behaviors and consequences in various 3D tissue constructs, as summarized in Figure 1.

| Construct Design | Model of Limited Diffusant out of Construct without Metabolism | Model of Unlimited Diffusant into Construct without Metabolism | Model of Unlimited Diffusant into Construct with Constant Metabolic Rate | Model of Limited Diffusant into Construct with Constant Metabolic Rate |
|---|---|---|---|---|
| 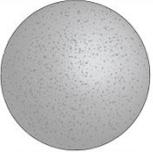 | Eq. (3) | Eq. (6) | Eq. (9) | Eq. (12) |
| 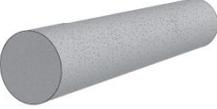 | Eq. (2) | Eq. (5) | Eq. (8) | Eq. (11) |
| 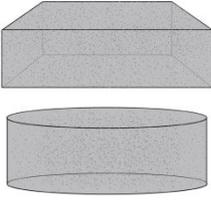 | Eq. (1) | Eq. (4) | Eq. (7) | Eq. (10) |

*Figure 1: Summary of Equations for Diffusion Modeling in 3D Tissue Constructs. Models of transient and steady-state diffusion in a variety of scenarios are provided, including for limited or unlimited diffusants, diffusion into or out of the tissue, with or without metabolism of the diffusant, and in constructs of slabs (1D – bottom row), cylinders (2D – middle row), or spheres (3D – top row).*

*2.2 The Influence of Diffusing Factors on Cell Signaling Pathways, Metabolism, and Potency State*

Although much remains to be learned regarding the role of diffusing nutrients in tissue development, it has recently become apparent that metabolic dynamics and nutrient supply can control epigenetic configurations of stem cells, and reciprocally, epigenetic networks control energetic processes and metabolic preferences in the cell. Moreover, these effects can play a significant role in stem cell potency, differentiation, and fundamental programmed processes in tissue development. Evidence suggests that the metabolic activity of a cell is actively modified by many of the same epigenetic reconfigurations that occur through changing states of stem cell potency and differentiation, and in general, while mature cells tend to favor the efficiency of oxidative metabolism, pluripotent stem cells (PSCs) tend to favor a glycolytic state [6]. Many other types of stem cells, including neural stem cells (NSCs), also favor anaerobic glycolytic metabolism [7,8] while mature neurons favor oxidative metabolism [9,10]. The oxidative preference of adult somatic cells is converted to a glycolytic preference early in reprogramming to a pluripotent state [11,12,13], and this may be related to the observation that glycolysis is also favored by malignantly-transformed cells, which resemble stem cells in their self-renewal and ability to endure hypoxic environments [14,15,16]. In cancer cells, this shift in metabolism from aerobic to anaerobic appears to be actively instigated by mutated genes like tumor suppressor protein p53 (*TP53*) and its downstream targets [17,18]. Similarly, in both embryonic and induced stem cells, these metabolic preferences are not merely passive consequences, but are in fact requisite in maintaining the pluripotent state and in reprogramming mature cells to a pluripotent state, and differentiation of pluripotent cells can be impeded unless these glycolytic processes switch to an oxidative metabolism [19,20,21].

Expression of several glycolytic enzymes is upregulated under conditions of hypoxia [16,22]. It is thought that the preferred glycolytic state of stem cells or cancer cells may serve to protect them from reactive oxygen species (ROS)—glycolysis enables the pentose phosphate pathway to produce the reduced form of nicotinamide adenine dinucleotide phosphate (NADPH), which keeps glutathione in a reduced state (GSH) for antioxidant protection [17]. Exposure of adult stem cells to ROS has been noted to prompt quiescent adult stem cells, including hematopoietic stem cells (HSCs), mesenchymal stem cells (MSCs), and neural stem cells (NSCs), out of quiescence and into proliferation [8], and exposure to ROS can also affect cell fate decisions [23,24]. Antioxidants like vitamin C can help reverse cell senescence and aid changes in epigenetic expression during reprogramming of induced pluripotent stem cells (iPSCs) [25]. The influence of a stem cell's energy status and mitochondrial function on its potency state may also explain why supplementation with electron-carrier coenzymes like $NAD^+$ precursor nicotinamide riboside has been found to rejuvenate neural and muscle stem cells [26]. In addition, despite less efficient energy production than oxidative phosphorylation, glycolysis and the pentose phosphate pathway also enable improved anabolic nutrient production for cell proliferation, including the synthesis of nucleotides, amino acids, and lipids [27].

Several sensors of energy usage and nutrient availability exist in the cell, including adenosine monophosphate-activated kinase (AMPK) and the mammalian target of rapamycin complex (mTORC). As adenosine triphosphate (ATP) energy is used, AMPK becomes activated by phosphorylation in order to inhibit anabolic processes that consume ATP and to activate catabolic processes that replenish ATP. Similarly, a low supply of amino acids will attenuate mTORC activity, which lowers anabolic activity in the cell. Under normoxic conditions, the 14-3-3 protein inhibits the tuberous sclerosis 2 protein (TSC-2), preventing expected inhibition of mTORC by the tuberous sclerosis protein 1/2 (TSC-1/2) complex. Hypoxia, however, upregulates REDD1 protein, which inhibits binding of the 14-3-3 protein to the TSC-1/2 complex, thereby enabling inhibition of mTORC [8,28]. Activated AMPK (whether from low glucose or oxygen supply or from high energy usage that exceeds energy supply) can inhibit mTORC activity, thereby inhibiting protein production. Interestingly, reprogramming to pluripotency decreases mTORC activity, likely due to the fact that Sox2 represses mTORC expression [29]. Moreover, activation of mTORC (e.g., by deletion of *TSC-2*) inhibits reprogramming to pluripotency, while inhibition of mTORC (e.g., with rapamycin) enhances reprogramming to a pluripotent state, and the inhibition of mTORC has also been shown to help maintain the population of stem cells and to suppress the production of ROS [8,30]. It should also be noted, however, that mTORC inhibition and AMPK activation can disrupt expression of Oct4, Sox2, or Nanog and drive differentiation of certain germ layer lineages [6,31,32]. Hypoxia can also lead to changes in histone deacetylase (HDAC) activity and histone phosphorylation via AMPK [33,34], both of which alter gene expression, as summarized in Figure 2.

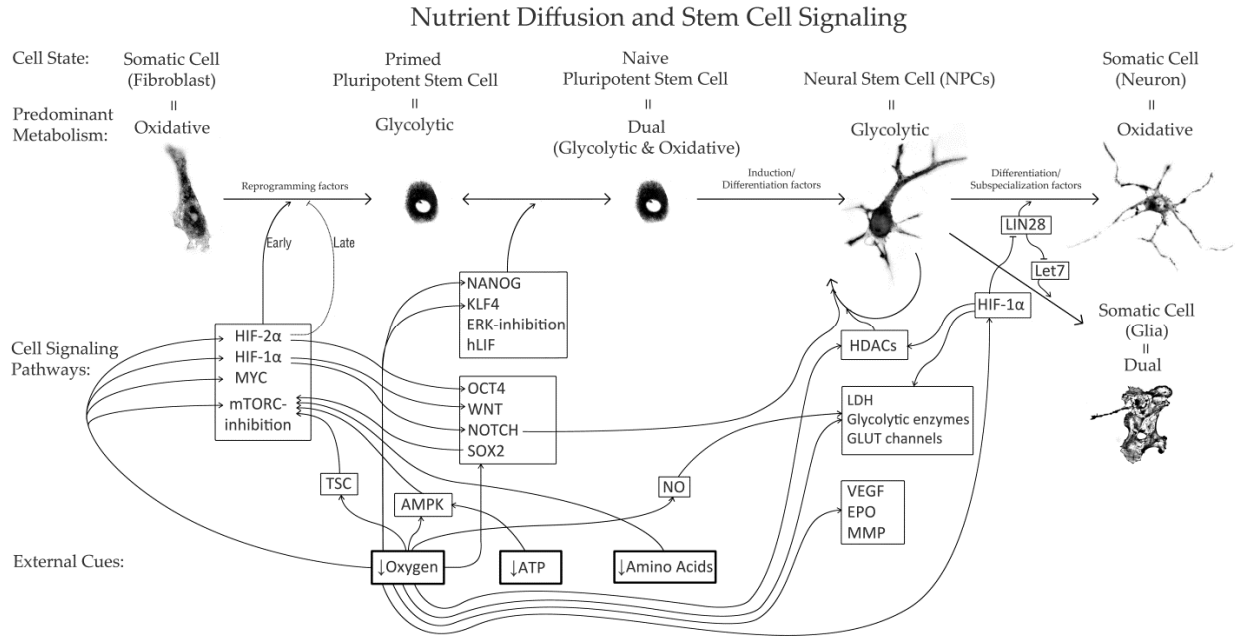

*Figure 2: Diagram of Nutrient Diffusion and Stem Cell Signaling. The effects of various nutrient diffusants involved in energy production and metabolism are shown to also influence stem cell states and cellular differentiation.*

The multipotency of neural stem/progenitor cells is also generally preserved by a hypoxic environment, including both physiologic hypoxia (2-3%) and more severe hypoxia that would likely otherwise threaten viability of mature neuronal cells [8,35,36]. Many types of endogenous stem cells are known to reside in a hypoxic niche, including hematopoietic stem cells (HSCs), mesenchymal stem cells (MSCs), and neural stem cells (NSCs), which preserves tissue-specific endogenous stem cell populations [37]. Metabolic processes are also tightly coupled with the balance of stem cell populations in many types of tissues, and forced over-activation of mTORC (e.g., by deletion or inhibition of mTORC-inhibitors or constitutive expression of mTORC-activators) is known to shorten and accelerate the cell cycle in NSCs and HSCs [8], which, depending on the context and conditions, can diminish the repopulation capacity of HSCs or expand the proliferative capacity of NSCs [30,38]. Also, Lin28, acting through the PI3K-mTORC pathway, has been shown to accelerate cell proliferation in mouse PSCs [39].

In human PSCs, much evidence suggests that hypoxia tends to promote and preserve the pluripotent state, both preventing differentiation and enhancing reprogramming efficiency [15,19,31,34,40]. In fact, hypoxia has been shown to induce expression of many of the same genes used as reprogramming factors for pluripotency, including *OCT4*, *SOX2*, *KLF4*, *NANOG*, *MYC*, and *LIN28* [15,41]. Because the choice of reprogramming factors substantially influences the quality and developmental potential of reprogrammed stem cells [42], environmental conditions of gasses, nutrients, and signaling factors will also influence these qualities, all of which are mediated by diffusion processes.

At greater distances from energy and nutrient sources where lower levels of nutrients will exist in the tissue due to diffusion limitations, cell pathways that favor pluripotent states are thus more likely to be active. This has an interesting correlate in cerebral organoids, where cortical neuron precursors migrate to and terminally differentiate at the external rim of the construct nearest the environmental oxygen supply, while deeper into the construct where oxygen is much more limited, neural stem cells may be better preserved and expand to supply future neural populations that fill the cortex. As organoid spheres expand, the hypoxic gradient can alter the position, timing, and fate of stem cells within the organoid and can also threaten cell viability. Diffusion modeling using Eq. 9 for oxygen & Eq. 12 for glucose has shown that central hypoxia in stem-cell-derived organoids is the main factor in limiting their maximal size and causing a central necrotic core if they grow beyond the limits of oxygen diffusion and metabolic consumption, though glucose could also become a limiting factor if feeding media is not replenished frequently enough [4].

Environmental availability of oxygen is known to regulate sets of hypoxia inducible factors (HIFs), and HIFs regulate the expression of many genes involved in stem cell state, cellular development, and metabolic functions. At normal atmospheric oxygen concentration and pressure, oxygen induces hydroxylation and ubiquitinization of HIFs by prolyl-hydroxylases (PHDs) and the von-Hippel-Lindau protein (pVHL), respectively, which then targets $\alpha$-subunits of HIFs for proteosomal degradation; with exposure to hypoxia, however, the $\alpha$-subunits of HIFs are stabilized and bind to their respective nuclear translocators (e.g., HIF1$\beta$ and HIF2$\beta$), where, in the nucleus, the HIFs then bind to various hypoxia-response elements for transcriptional regulation [43]. Similarly, enzymes like JmjC histone lysine demethylase (KDMs) are sensitive to specific oxygen concentrations and influence epigenetic regulation of the cell [44]. Both HIF1$\alpha$ and HIF2$\alpha$ are required for reprogramming to pluripotency, and the activity of either one alone activates the accompanying metabolic change to anaerobic glycolysis, although HIF2$\alpha$ activity in the late stages of reprogramming can inhibit the reprogramming process [12].

Although ESCs and iPSCs are both pluripotent stem cells with equivalent functional potential and only minor epigenetic differences between them [45], further sub-states of pluripotency have emerged, including the concept of naïve versus primed pluripotent states. The naïve state tends to prefer oxidative metabolism (but utilizes both glycolytic and oxidative phosphorylation) and seems to represent the earliest state of embryonic development before implantation into the womb, while the primed pluripotent state favors glycolytic energy production and represents a more mature post-implantation state where DNA methylation patterns have already undergone significant changes [46,47]. Pluripotent cells can be coaxed into either state with various intrinsic and extrinsic factors [48,49]. Among the factors that promote a naïve state are the expression of Nanog and Klf4, both of which are promoted by hypoxia (Figure 2), again suggesting that a low oxygen environment likely favors the naïve pluripotent state, although it is not known if this alone can be sufficient to induce or maintain such a state in certain cells.

Hypoxia also activates other genes associated with stem cell states and cell development, including expression of Notch, WNT, and SHH, all via HIF1α [50,51] (Figure 2). It is not yet clear whether the mechanism of modulation of some hypoxia-responsive genes (like *MYC*, *NANOG*, or *KLF4*) is mediated directly by HIFs or other signaling pathways, but evidence suggests that some of these genes are regulated at least in part by non-coding microRNAs (miRNA). For example, *MYC* can be regulated by miRNA-210, which itself is controlled by both HIF1α and 2α [52], and Lin28 RNA-binding proteins inhibit let-7 miRNAs (which normally act as tumor suppressors), with the result that Lin28 and let-7 act as mutually antagonistic regulators of several downstream processes including glycolytic metabolism and cell growth via an mTORC-dependent pathway [53].

Several of the above factors are also involved in the early development of regional identity of neural tissue, including SHH, which generally drives ventralization, and WNT, which generally drives caudalization and dorsalization [5]. WNT and SHH activities are both increased under hypoxia via HIF1α, and both the WNT and SHH signaling pathways act through the β-catenin transducer, which activates effectors like FGF and Noggin to inhibit SMAD signaling pathways, as shown in Figure 3. SMAD inhibition is often used in directing differentiation of stem cells to a neuronal fate, which is typically done in culture with small molecule inhibitors of BMP and TGF-β that result in SMAD4-inhibition [54]. BMP and TGF-β can also activate Notch via SMAD1, as can hypoxia itself via HIF1α-mediated signals [55], and TGF-β and SMAD4 can also exhibit reciprocal inhibition with MYC [56].

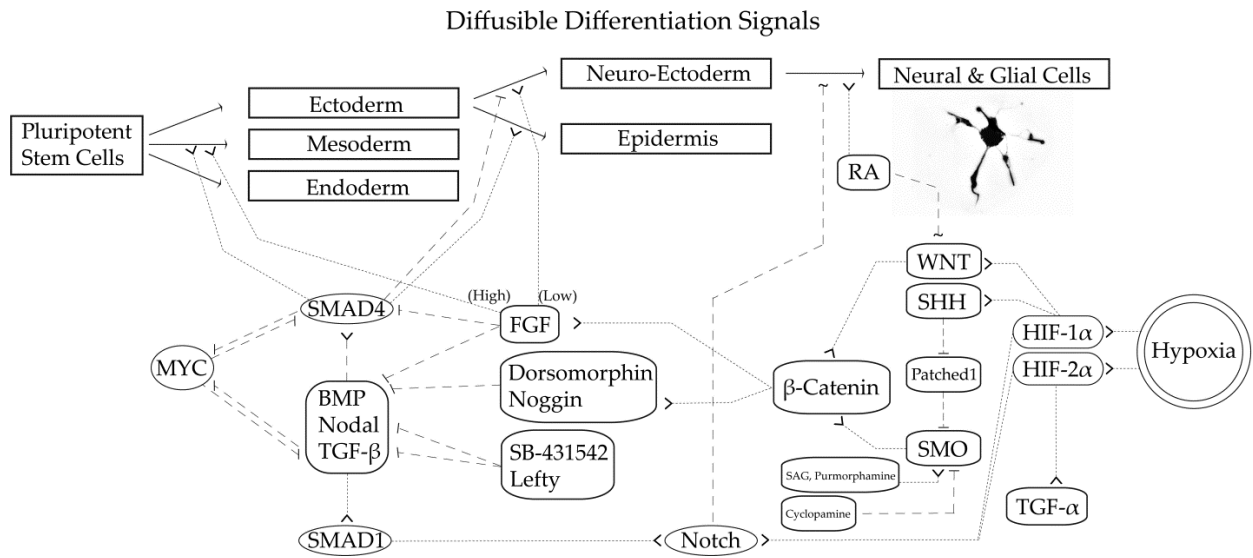

*Figure 3: Diagram of Diffusible Stem Cell Differentiation Signals. An overview is shown for how several signaling pathways can be used to guide differentiation from pluripotency to neuroglial lineage. Interestingly, many of the factors used in directed differentiation modulate the same molecular pathways used in oxygen signaling.*

Further evidence also shows that specific oxygen levels can have differential effects on germ layer specification and subsequent cell fate decisions, including whether neural progenitors progress to neurons or glia, depending on conditions and context that remain to be fully defined [41,57,58]. Exposure of spheroid cultures of human PSCs to transient hypoxia (2% vs 21% $O_2$), for example, was shown to drive a neural lineage over that of cardiovascular or musculoskeletal lineages and to change neuronal cell fate to a glial cell fate by acting through HIF1$\alpha$ to inhibit Lin28 expression [41]. Other work, however, has shown that inhibition of mTORC limits pluripotency and proliferation while enhancing differentiation of human PSCs towards endodermal and mesodermal lineages [32]. Differentiation of neural crest stem cells (NCSCs) under hypoxic conditions (1-5% $O_2$) has been shown to greatly expand and alter the possible array of mature neural subtypes, and activation of Notch signaling can promote maintenance of the neural precursor population and fate [55]. Physiologic hypoxia (2-3% $O_2$) has also been found to promote neurogenesis while severe hypoxia (<1% $O_2$) impeded both neurogenesis and gliogenesis [59].

Nitric oxide (NO) is another diffusible gas in tissues, which in the brain is produced by neural cells, glial cells, and vascular endothelial cells, both by constitutive and inducible mechanisms, particularly in response to ischemia in stroke, septic shock, or other poor perfusion events via various NO-synthase (NOS) isoforms as well as by other NOS-independent mechanisms [60,61]. NO can inhibit mitochondrial respiration and upregulate glycolysis and shunting through the pentose phosphate pathway, which may provide neuroprotection against free radical toxicity, mitochondrial damage, and apoptosis [61,62]. However, the production of NO in response to hypoxia or ischemia appears to depend on the particular conditions and timing of oxygen deprivation and reperfusion, and NO may play either protective or toxic roles depending on local conditions and tissue cell types, with neuroglial-produced NO playing a possible neurotoxic role while endothelial-produced NO may play a protective role by enhancing vascular perfusion. Consequently, both NO and NO-suppression have been suggested as therapeutic interventions for ischemic conditions in tissues of the brain and heart. Similarly, another gas affecting stem cell survival and differentiation is carbon dioxide ($CO_2$), with evidence that higher levels of $CO_2$ (10% versus the standard 5%) enabled the formation of larger and higher-quality neurospheres, induced significantly greater Nestin, Pax6, Sox2, and Foxg1 expression in neuroprogenitor populations, and facilitated the genesis of glutamatergic, cholinergic, dopaminergic, and GABAergic neuronal subtypes [63]. Importantly, these data provide further evidence that cell populations in 3D tissue cultures are influenced not just by inward diffusion of outside nutrients, but also by factors diffusing outwardly from inside the tissue construct.

Local variations in gas and nutrients within a tissue construct can therefore induce significant variations in cell and tissue identity, although it remains unknown to what extent the microenvironment of diffusant substances can regulate and guide essential developmental processes and cell states in the many different tissues of the human body. The effects of hypoxia on differentiation likely depend substantially not just on the exact amount of ambient oxygen or nutrient but also on genetic networks that are active or inactive in the cell and the phase of differentiation and cell maturity. A better understanding of oxygen concentrations within 3D cell cultures along with careful analysis of intermediate states of differentiation and development will thus help delineate the interactions of complex signaling cues and help resolve conflicting results.

*2.3 The Influence of Nutrient Signaling and Metabolism on Neurological Disease*

Several factors involved in hypoxia signaling pathways are also known to play critical roles in neurological development and disease. For example, the tuberous sclerosis protein (TSC) complex, which acts as a tumor suppressor and, when mutated, causes the condition of tuberous sclerosis, acts to inhibit the mTOR complex, but is itself normally inhibited under normoxia. Subependymal nodules and giant cell astrocytomas associated with tuberous sclerosis are thought to be due to abnormal proliferation and migration of neural stem/progenitor cells (NSPCs) [64], and recent clinical trials show these types of tumors, and potentially others like glioblastoma and medulloblastoma, may be treated with mTORC-inhibitors like sirolimus, everolimus, dactolisib, XL765, or INK128 [65,66]. Similarly, neurofibromatosis 1 is caused by disruption of the Ras-effector pathway which in turn can act through the TSC complex to exert its effects on mTORC, and the optic gliomas associated with this condition may also benefit from mTORC-inhibitors [67]. Moreover, mutations in pVHL cause von Hippel Lindau syndrome, with a constellation of tumors, vascular malformations, and cysts. Mutations or disruptions of SHH, or its downstream effector Patched1, can result in holoprosencephaly and craniofacial abnormalities. Metabolic impairments of mitochondrial function and susceptibility to oxidative stress are also suspected to play a major role in the pathogenesis of Parkinson's diseases via mutant LRRK2, PARK2, PINK1, or Miro proteins [68].

Hypoxia is also known to inhibit numerous other signaling factors relevant to vascular development, cell migration, and tumor growth, including VEGF, EPO, and matrix metalloproteinase activity [51,69,70], and, as is seen in Figures 2 and 3, many of these pathways overlap and intersect with each other. The mechanisms whereby the mitochondrial, nuclear, and cytoplasmic signaling work together (including causes and effects) to dictate metabolic preferences, energy production, oxidative protection, potency preservation, and cell fates still remains to be elucidated. Because metabolic functions and mitochondrial activity play essential roles in maintenance of and differentiation from stem cell states [71] researchers are obliged to be aware of difficulties that may arise in accurately modeling or restoring metabolic and mitochondrial diseases with patient-affected iPSCs compared with normal iPSCs.

Pathological hypoxia in development is linked to several intractable neurological conditions, including encephalopathy, epilepsy, and cognitive impairments. Interestingly, however, exposure of many types of adult cells, including neurons and cardiomyocytes, to sub-lethal intervals of hypoxia can also enable cells to survive subsequent insults of more severe hypoxia, a phenomenon known as preconditioning, which may have implications for protecting organs and tissue from hypoxic-ischemic events [72]. The proper contexts and molecular mechanisms of this protective conditioning effect are still being fully elucidated [73], but this phenomenon may be useful for helping therapeutic cells and synthetic tissues survive after implantation into the body since the survival of these implanted tissues prior to vascularized integration is dependent on diffusion of nutrients from surrounding tissues. The influence of gas and nutrient concentrations on gene expression and cell signaling pathways will also have many consequences for 3D tissue culture protocols and media formulations.

Altogether, the collection of evidence shows that diffusant signaling is a crucial factor in all stem cell function, particularly for neural tissue development, and therefore an understanding of diffusion properties and profiles in tissue is essential to the study of tissue development. The ability of specific levels of gas, nutrients, and signaling factors to influence states of potency and differentiation means that diffusion can have significant effects on the composition, shape, and function of various tissues throughout the body. The rate and timing of stem cell self-renewal, quiescence, and differentiation will tightly influence the size and capacity of the endogenous stem cell population in adult tissues and will also influence the quantity, balance, structure, and function of more specific cell identities in synthetic tissues. The mechanisms that regulate and carry out these functions, however, are not yet well understood despite the fact that they exert tremendous influence over numerous developmental and physiological processes.

## 3. MODELS OF DIFFUSION IN TISSUE CONSTRUCTS

The represented constants and variables for the following models are listed in the nomenclature summary (Table 1). In order to enable exact mapping of physical nutrient gradients through a 3D tissue construct at any given point in time, diffusion models were derived from the diffusion equation given by Fick's laws, as applied to tissue constructs in the shape of a rectangular slab (s=1), a cylinder (s=2), or a sphere (s=3), and with a homogenous metabolic consumption rate of $\varphi$:

$$\frac{\partial C}{\partial t} - \varphi = \frac{1}{r^{s-1}} \frac{\partial}{\partial r}\left(r^{s-1} D \frac{\partial C}{\partial r}\right)$$

Analytic solutions describing diffusion of substances into 3D tissue constructs were recently described by the author, including novel models for diffusant substances that are metabolized by cells in the tissue construct (e.g., nutrients like glucose and oxygen) and for diffusants that are not metabolized by cells (e.g., certain cell signaling factors) as well as for both limited and unlimited diffusants (like glucose and oxygen, respectively), as summarized in Table 2. Approaches for finding various solutions to the diffusion equation and creating applicable models have been described by many authors [74,75,76,77], which is discussed in more detail in reference [4]. Equations describing outward diffusion from a tissue construct (e.g., for diffusing factors that are produced by or embedded in the tissue construct) can be derived in a similar manner with altered boundary conditions, the solutions of which are also presented in Table 2 as Eqs. 1-3. Typical values for diffusion parameters can be found in Table 3, and complete descriptions of boundary conditions and parameters for operating these models can be found in Table 5 and in reference [4].

Because of difficulty in measuring gas or nutrient gradients through small, metabolically-active spheres, mathematical models provide a means for estimating diffusion gradients based on known diffusion mechanics and a few basic assumptions. The models are valid under the assumptions that cells are homogenously distributed and diffusivity is isotropic in the tissue construct, that diffusion occurs primarily along a single axis through a slab (1D case) or along the radial axis of a cylinder or sphere (2D or 3D cases), and, for the metabolic cases, that consumption rates of the diffusant (represented by $\varphi$) are constant in the construct during each modeled time interval. In the case of slab constructs, the surface of the construct is at $x = 0$ and the thickness of the construct is represented by $x = T$, whereas the surface of the radial constructs is at $r = R$ and the center at $r = 0$, and solutions to the models are valid within the domain of $0 < x < T_{max}$ or $0 < r < R_{max}$. In models of outward diffusion, $C_o$ is the concentration of substance that diffuses from the center of the tissue construct, and in models of inward diffusion, $C_o$ is the initial concentration of substance at the outer edge of the tissue construct.

In most cases, the initial concentration in a tissue construct ($C_i$) is zero, such as when newly-formed organoids are introduced into a media environment of glucose, oxygen, or other molecular factors, but in cases where the diffusant substance of interest is already present in the biomaterial, its initial concentration ($C_i$) is subtracted from the driving concentration of $C_o$ and also subtracted from the total concentration $C$. In other words, "$C_o$" is replaced with "$(C_o - C_i)$") and "$C$" is replaced with "$(C - C_i)$". In the cases of diffusion out of a tissue construct, it is assumed for simplicity that the diffusant is diluted to negligible levels once it leaves the tissue construct (e.g., as when the tissue construct volume is small relative to the surrounding media). If this assumption is not made, the diffusion into the media must also be accounted for and the total initial amount of diffusant must be spread over the total volume of the construct and the media at steady state, and the release of diffusant from the construct will influence its media concentration and by extension also influence the diffusive driving force over time.

These analytic models thus allow mapping of concentration gradients through space and time for a variety of molecular signaling factors and nutrients into or out of 3D tissues in the shape of slabs, cylinders, or spheres. Of course, many more complex conditions and systems will require more complicated mathematical models that necessitate numerical approximations with computational software rather than complete analytic solutions, but this serves as an important foundation for mathematical descriptions of diffusion in most general cases of developing tissues. Finally, it is also important to note that actual gradient concentrations could deviate from predicted values with varying rates of metabolism, with varying inhomogeneities in the materials or cell densities, or with other pertinent forces on the tissue construct or diffusant molecules.

| Nomenclature Summary | | | |
|---|---|---|---|
| $C$ | Concentration | $\pi$ | 3.14159…. |
| $C_o$ | Initial concentration of diffusant at interface of source and tissue construct | $\varphi$ | Metabolic consumption rate for tissue construct (in units of mol/Ls, i.e., $\varphi = m\rho$) |
| $\bar{C}$ | Average concentration in tissue construct | $R$ | Outer radius of a radial tissue construct |
| $C_i$ | Initial baseline concentration in tissue construct | $R_{max}$ | Maximal radius of a radial tissue construct |
| $C_{critical}$ | Critical concentration | $R_{max_t}$ | Maximal radius as a function of time |
| $C_{media}$ | Media concentration | $r$ | Radial distance or spatial position |
| $D$ | Diffusion coefficient | $\rho$ | Density of cells in tissue construct |
| $d$ | Ordinary Derivative | $s$ | Spatial dimension of system |
| $\partial$ | Partial Derivative | $\Sigma$ | Summation series in sigma notation |
| erf | Error function | $t$ | Time |
| erfc | Complementary error function | $T$ | Thickness of linear tissue construct |
| $e$ | 2.71828…. | $V_c$ | Volume of tissue construct |
| $J_q()$ | Bessel function of the first kind and order $q$ | $V_m$ | Volume of media around tissue construct |
| $\lambda_n$ | Eigenvalues | $x$ | Spatial position or linear depth into construct |
| $m$ | Metabolic rate per cell | $y$ | Spatial position in Cartesian coordinates |
| $n$ | Index term in sigma summation series | $z$ | Spatial position in Cartesian coordinates |

*Table 1: Nomenclature Summary of Variables, Functions, & Constant Parameters.*

Diffusion of limited diffusant out from tissue construct (no metabolism of diffusant)

1D (Eq. 1)
$$C(x,t) = C_o \left[ \text{erf}\left(\frac{x}{\sqrt{4Dt}}\right) + \sum_{n=1}^{\infty} \left\{ \text{erf}\left(\frac{(2n)T-x}{\sqrt{4Dt}}\right) - \text{erf}\left(\frac{(2n)T+x}{\sqrt{4Dt}}\right) \right\} \right]$$

2D (Eq. 2)
$$C(r,t) = \frac{2C_o}{R} \sum_{n=1}^{\infty} e^{-(\lambda_n)^2 Dt} \frac{J_o(r\lambda_n)}{\lambda_n J_1(R\lambda_n)}$$ where $\lambda_n$ are found from Table 4.

3D (Eq. 3)
$$C(r,t) = \frac{2RC_o}{\pi r} \left[ \sum_{n=1}^{\infty} \frac{-1^{(n+1)}}{n} e^{-\left(\frac{n\pi}{R}\right)^2 Dt} \sin\left(\frac{n\pi r}{R}\right) \right]$$

Diffusion of unlimited diffusant into tissue construct (no metabolism of diffusant)

1D (Eq. 4)
$$C(x,t) = C_o \left[ \text{erfc}\left(\frac{x}{\sqrt{4Dt}}\right) + \sum_{n=1}^{\infty} (-1)^{n+1} \left\{ \text{erf}\left(\frac{(2n)T-x}{\sqrt{4Dt}}\right) - \text{erf}\left(\frac{(2n)T+x}{\sqrt{4Dt}}\right) \right\} \right]$$

2D (Eq. 5)
$$C(r,t) = C_o \left[ 1 - \frac{2}{R} \sum_{n=1}^{\infty} e^{-(\lambda_n)^2 Dt} \frac{J_o(r\lambda_n)}{\lambda_n J_1(R\lambda_n)} \right]$$ where $\lambda_n$ are found from Table 4.

3D (Eq. 6)
$$C(r,t) = C_o + \frac{2RC_o}{\pi r} \left[ \sum_{n=1}^{\infty} \frac{-1^n}{n} e^{-\left(\frac{n\pi}{R}\right)^2 Dt} \sin\left(\frac{n\pi r}{R}\right) \right]$$

Diffusion of unlimited diffusant into tissue construct (with zero-order metabolism)

1D (Eq. 7)
$$C(x,t) = C_o + \frac{\varphi x^2}{2D} - \frac{\varphi T x}{D} - \frac{2C_o}{\pi} \sum_{n=1}^{\infty} \frac{1}{n} e^{-\left(\frac{n\pi}{T}\right)^2 Dt} \sin\left(n\pi \varphi x \frac{2T-x}{2C_o D}\right)$$

2D (Eq. 8)
$$C(r,t) = \frac{C_o r^2}{R^2} + \frac{2C_o}{\pi} \left[ \sum_{n=1}^{\infty} \frac{-1^n}{n} e^{-\left(\frac{n\pi}{R}\right)^2 Dt} \sin\left(n\pi \frac{r^2}{R^2}\right) \right]$$ where $R_{max} = \frac{\varphi r^2}{4D}$

3D (Eq. 9)
$$C(r,t) = \frac{C_o r^2}{R^2} + \frac{2C_o}{\pi} \left[ \sum_{n=1}^{\infty} \frac{-1^n}{n} e^{-\left(\frac{n\pi}{R}\right)^2 Dt} \sin\left(n\pi \frac{r^2}{R^2}\right) \right]$$ where $R_{max} = \frac{\varphi r^2}{6D}$

Diffusion of limited diffusant into tissue construct (with zero-order metabolism)

1D (Eq. 10)
$$C(x,t) =$$

$$\left(C_o - \varphi t \frac{V_c}{V_m} - \bar{C}\frac{V_c}{(V_m + V_c)}\right) + \frac{\varphi x^2}{2D} - x\sqrt{\frac{2\varphi\left(C_o - \varphi t \frac{V_c}{V_m} - \bar{C}\frac{V_c}{(V_m + V_c)}\right)}{D}}$$

$$-\frac{2\left(C_o - \varphi t \frac{V_c}{V_m} - \bar{C}\frac{V_c}{(V_m + V_c)}\right)}{\pi}\sum_{n=1}^{\infty}\frac{1}{n}e^{-\left(\frac{n\pi}{T}\right)^2 Dt}\sin\left(n\pi x\frac{\sqrt{8D\varphi\left(C_o - \varphi t \frac{V_c}{V_m} - \bar{C}\frac{V_c}{(V_m + V_c)}\right)} - \varphi x}{2D\left(C_o - \varphi t \frac{V_c}{V_m} - \bar{C}\frac{V_c}{(V_m + V_c)}\right)}\right)$$

where $\bar{C} = C_o\left[1 - \frac{8}{\pi^2}\sum_{n=1}^{\infty}\frac{1}{(2n-1)^2}e^{-\left(\frac{(2n-1)\pi}{2T}\right)^2 Dt}\right]$ and $T_{max_t} = \sqrt{\frac{2sD\left(C_o - \varphi t \frac{V_c}{V_m} - \bar{C}\frac{V_c}{(V_m + V_c)}\right)}{\varphi}}$

2D    (Eq. 11)
$$C(r,t) = \frac{\varphi(r - R_{max} + R_{max_t})^2}{4D}$$
$$+ \frac{2\left(C_o - \varphi t \frac{V_c}{V_m} - \bar{C}\frac{V_c}{(V_m + V_c)}\right)}{\pi}\left[\sum_{n=1}^{\infty}\frac{-1^n}{n}e^{\left(\frac{-\varphi t (n\pi)^2}{\left(4C_o - 4\varphi t \frac{V_c}{V_m} - \bar{C}\frac{4V_c}{(V_m + V_c)}\right)}\right)}\sin\left(\frac{\varphi n\pi(r - R_{max} + R_{max_t})^2}{4D\left(C_o - \varphi t \frac{V_c}{V_m} - \bar{C}\frac{V_c}{(V_m + V_c)}\right)}\right)\right]$$

where $\bar{C} = C_o\left[1 - \sum_{n=1}^{\infty}\frac{4}{(\lambda_n R)^2}e^{-\lambda_n^2 Dt}\right]$ and $R_{max_t} = \sqrt{\frac{4D\left(C_o - \varphi t \frac{V_c}{V_m} - \bar{C}\frac{V_c}{(V_m + V_c)}\right)}{\varphi}}$

3D    (Eq. 12)
$$C(r,t) =$$
$$\frac{\varphi(r - R_{max} + R_{max_t})^2}{6D} + \frac{2\left(C_o - \varphi t \frac{V_c}{V_m} - \bar{C}\frac{V_c}{(V_m + V_c)}\right)}{\pi}\left[\sum_{n=1}^{\infty}\frac{-1^n}{n}e^{\left(\frac{-\varphi t (n\pi)^2}{\left(6C_o - 6\varphi t \frac{V_c}{V_m} - \bar{C}\frac{6V_c}{(V_m + V_c)}\right)}\right)}\sin\left(\frac{\varphi n\pi(r - R_{max} + R_{max_t})^2}{6D\left(C_o - \varphi t \frac{V_c}{V_m} - \bar{C}\frac{V_c}{(V_m + V_c)}\right)}\right)\right]$$

where $\bar{C} = C_o\left[1 - \frac{6}{\pi^2}\sum_{n=1}^{\infty}\frac{1}{n^2}e^{-\left(\frac{n\pi}{R}\right)^2 Dt}\right]$ and $R_{max_t} = \sqrt{\frac{6D\left(C_o - \varphi t \frac{V_c}{V_m} - \bar{C}\frac{V_c}{(V_m + V_c)}\right)}{\varphi}}$

*Table 2: Analytic Models of Diffusion. Twelve different equations are given for modeling several different scenarios of diffusion of nutrients and signaling factors in 3D tissue constructs.*

| Typical Ranges of Diffusant Parameters in Tissue Systems | |
|---|---|
| Diffusion Coefficient ($D$) | $10^{-8}$ to $10^{-11}$ m$^2$/s |
| Initial Concentration ($C$) | 0.01 to 100 mM |
| Metabolic Rate of Cell ($m$) | $10^{-15}$ to $10^{-18}$ mol/s |

*Table 3: Parameter Ranges for Diffusing Molecules in Tissue Systems. Typical ranges of important parameters in tissue diffusion systems are provided [4]. These parameters can be used to predict the maximal viable size of a tissue construct based on limitations of diffusing nutrients, where the maximal depth of a tissue construct is $\sqrt{\frac{2sC_oD}{\varphi}}$ and where the metabolic rate of a tissue construct ($\varphi$) can be determined from the average density of cells ($\rho$) multiplied by the average metabolic rate of the cells ($m$).*

| Bessel Function Roots | | |
|---|---|---|
| | $J_o(x)$ (n$^{th}$ roots) | $\lambda_n$ for $J_o(\lambda_n R)$ |
| n=1 | 2.4048 | 2.4048/R |
| n=2 | 5.5201 | 5.5201/R |
| n=3 | 8.6537 | 8.6537/R |
| n=4 | 11.7915 | 11.7915/R |
| n=5 | 14.9309 | 14.9309/R |
| n=6 | 18.0711 | 18.0711/R |
| n=7 | 21.2116 | 21.2116/R |
| n=8 | 24.3525 | 24.3525/R |
| n=9 | 27.4935 | 27.4935/R |
| n=10 | 30.6346 | 30.6346/R |

Table 4: Bessel Function Roots. The first ten roots of the first kind of Bessel function of order zero, used for cylindrical diffusion solutions.

Eq. 1
$$C(x,0) = C_o \text{ for } 0 \leq x \leq T$$
$$C(0,t) = 0 \text{ for } t > 0$$
$$\partial C(\infty,t)/\partial x = 0$$

Eq. 2
$$C(r,0) = C_o \text{ for } 0 \leq r \leq R$$
$$C(R,t) = 0 \text{ for } t > 0$$
$$\partial C(0,t)/\partial r = 0$$

Eq. 3
$$C(r,0) = C_o \text{ for } 0 \leq r \leq R$$
$$C(R,t) = 0 \text{ for } t > 0$$
$$\partial C(0,t)/\partial r = 0$$
$$\lim_{r \to 0} C(r,t) = bounded$$

Eq. 4
$$C(x,0) = 0 \text{ for } 0 \leq x \leq T$$
$$C(0,t) = C_o \text{ for } t > 0$$
$$\partial C(\infty,t)/\partial x = 0$$

Eq. 5
$$C(r,0) = 0 \text{ for } 0 \leq r \leq R$$
$$C(R,t) = C_o \text{ for } t > 0$$
$$\partial C(0,t)/\partial r = 0$$

Eq. 6
$$C(r,0) = 0 \text{ for } 0 \leq r \leq R$$
$$C(R,t) = C_o \text{ for } t > 0$$
$$\partial C(0,t)/\partial r = 0$$
$$\lim_{r \to 0} C(r,t) = bounded$$

Eq. 7
$$C(x,0) = 0 \text{ for } 0 \leq x \leq T$$
$$C(T,t) = 0$$
$$C(0,t) = C_o$$

Eq. 8
$$C(r,0) = 0 \text{ for } 0 \leq r \leq R$$
$$C(R,t) = C_o$$
$$C(0,t) = 0$$

Eq. 9
$$C(r,0) = 0 \text{ for } 0 \leq r \leq R$$

$$C(R,t) = C_o$$
$$C(0,t) = 0$$

Eq. 10

$$C(x,0) = 0 \text{ for } 0 \leq x \leq T$$
$$C(T,t) = 0$$
$$C(0,t) = C_o - \varphi t \frac{V_c}{V_m} - \bar{C} \frac{V_c}{(V_m + V_c)}$$

Eq. 11

$$C(r,0) = 0 \text{ for } 0 \leq r \leq R$$
$$C(R,t) = C_o - \varphi t \frac{V_c}{V_m} - \bar{C} \frac{V_c}{(V_m + V_c)}$$
$$C(0,t) = 0$$

Eq. 12

$$C(r,0) = 0 \text{ for } 0 \leq r \leq R$$
$$C(R,t) = C_o - \varphi t \frac{V_c}{V_m} - \bar{C} \frac{V_c}{(V_m + V_c)}$$
$$C(0,t) = 0$$

*Table 5: Diffusion Model Boundary Conditions. The initial and boundary conditions of each respective equation are provided.*

## 4. CONCLUSIONS

With advancing abilities to create tissue and organoid structures from stem cells and biomaterials, and with new and ongoing discoveries of how gas, nutrient, and signaling factor concentrations produce differential effects on stem cell state and function, a novel role of diffusion modeling has emerged for study of stem cell functions and developmental processes in three-dimensional (3D) tissues. Ideal compositions for formation of targeted tissue structures, cellular identities, and functional neural networks remain to be explored, and in the course of these endeavors it is important to recognize that the ability to properly guide differentiation of stem cells and development of tissues for therapeutic use is strongly dependent on how gasses, nutrients, and signaling factors diffuse in cultured tissue constructs. In fact, growth and development of cellular tissues both are influenced by and influence the internal diffusion dynamics as a reciprocal interaction. This work describes many complex interactions of diffusant substances in stem cell biology and presents several unique analytic models for understanding diffusion phenomena in tissue constructs, thereby enabling modeling of oxygen and nutrient delivery to cells and study of mass transport and spatial gradients that form in 3D tissue constructs under a variety of conditions. The use of engineered combinations of cells, biomaterials, and biochemical diffusing factors is likely to one day enable guided differentiation and detailed control of cellular organization in synthetic tissue constructs. These concepts and investigations will therefore have significant impact on regenerative approaches for many otherwise disparate diseases and injuries, particularly those of the nervous system, including stroke, spinal cord injury, cancer, neurodegenerative diseases, and many other neurogenetic syndromes and developmental abnormalities.

## SUPPLEMENTAL INFORMATION

A Matlab script with a simple graphical user interface is provided as supplementary material at *http://dx.doi.org/10.1089/scd.2017.0066* which can solve the diffusion models in tissue constructs by simply inputting appropriate parameters of the system. Instructions for operating the files are also provided.

## AUTHOR DISCLOSURE STATEMENT

The author declares no potential conflicts of interest with respect to the research, authorship, or publication of this article. No grant funding was provided for this work.